\documentclass[graybox, envcountchap]{svmult}

\usepackage{mathptmx}        
\usepackage{amsmath}
\usepackage{aastexx}
\usepackage{amssymb}
\usepackage{color}
\usepackage{helvet}          
\usepackage{courier}         
\usepackage{dirtree}

\usepackage{makeidx}        
\usepackage{graphicx}        
\usepackage{subfig}
\usepackage{multicol}        
\usepackage[bottom]{footmisc}

\usepackage{hyperref}        
\hypersetup{colorlinks=true,urlcolor=blue}

\usepackage[misc]{ifsym}

\makeindex             

\newcommand{\surfbrightness}{421 }

\begin{document}


\title{The Mira Distance Ladder}
\author{Caroline D. Huang}
\institute{Caroline D. Huang (\Letter) \at Center for Astrophysics | Harvard \& Smithsonian, 60 Garden St., Cambridge, MA 02138 \email{caroline.huang@cfa.harvard.edu}}
%
%
\maketitle

\abstract{Here we provide a review of Mira variables, their basic properties, and Period-Luminosity Relations with an emphasis on their role in measuring the Hubble Constant. The usage of multiple independent distance indicators and methods is crucial to cross-checking systematic uncertainties in distance measurements and in reinforcing previous findings of the Hubble tension. To this end, Mira variables serve as an alternative Type Ia Supernova calibrator to the more commonly-used Cepheid variables or Tip of the Red Giant Branch method. They also have the potential to expand the number of local SN Ia calibrators by calibrating previously-inaccessible SNe Ia. Short-period ($ P \lesssim 400 $ d) O-rich Miras are a ubiquitous older population that can reach galaxies not hosting the younger Cepheids variables or out of reach to the old but fainter Tip of the Red Giant Branch. With the current and upcoming focus on infrared observations, Miras, which can be discovered and characterized using exclusively near-infrared and infrared observations, will be particularly useful in obtaining distances to astrophysical objects. Long-period Miras ($P \gtrsim$ 400 d) are highly luminous variables that  have the potential to measure H$_0$ directly, excluding Type Ia SNe altogether in the distance ladder. }


\section{Introduction}
\label{sec:intro} 

Mira variables are fundamental-mode, radially-pulsating tip of the Asymptotic Giant Branch (AGB) stars with periods ranging from $\sim 100 - 1000$ days or longer. They were the first class of periodic variables to be identified, starting with the discovery of Mira (\textit{Omicron Ceti}) by David Fabricius on August 3, 1596. With their high luminosities and the largest amplitudes of any regular pulsator ($> 2.5$ $m_V$ \cite{Samus_2017}), they remain a popular and easy target for amateur astronomers to this day. 

As highly-evolved AGB stars, they contribute to the chemical enrichment of the interstellar medium (ISM) through stellar mass loss \cite{Boyer_2015a} and are also extremely luminous --- generally the brightest stars in an intermediate-to-old stellar population. They also also ubiquitous, as nearly all stars ($0.8 M_\odot < M < 8 M_\odot$) will experience a Mira phase in evolution. 

Period-luminosity relations (P-L relations) for Miras have been known for several decades and have been employed to measure distances to many Local Group galaxies (\cite{Feast_1989} and references therein). However, the persistent Hubble tension has resulted in renewed interest in Mira distances as an alternative and cross-check to the more commonly-used Cepheid variables and Tip of the Red Giant Branch (TRGB) thanks to their age (intermediate-age Miras are $\sim$ 3 Gyr) and luminosity (up to a few $10^4 L_{\odot}$). With their low effective temperatures ($T_{\rm eff} < 3500$\ K) they are particularly attractive in light of the upcoming decade's focus on NIR and IR wavelengths since the spectral intensity of Miras with thin circumstellar shells (such that those currently used in distance measurements) peaks between 1-2$\mu$m. 

Miras may be classified into two spectral types -- Carbon- (C-) and Oxygen- (O-) rich. Current distance measurements using Miras employ exclusively short-period ($P \lesssim 400$ d), O-rich variables, which have been shown to have a tight ($\sigma \sim 0.12$ mag) P-L relation in the near-infrared. C-rich Miras have larger scatter at shorter wavelengths, and longer-period ($P \gtrsim 400$ days) Miras may follow a different P-L relation due to hot-bottom burning \cite{Whitelock_2003}. 

In this review, we will briefly discuss the  evolutionary status of Miras in Section \ref{sec:evolution}. Next, we will give an overview of the history of various Mira Period-Luminosity Relations and some local measurements and studies with Miras in Section \ref{sec:PLRs}. In Section \ref{sec:H0} we will give outline the previous efforts related to measuring H$_0$ with Miras. In Section \ref{sec:longperiod} we will briefly discuss prospects and previous observations of long-period Miras.
In Section \ref{sec:comparison} we compare Miras with other stellar primary distance indicators and conclude in Section \ref{sec:conclusions}. 

\begin{figure}
\centering
\includegraphics[scale=0.6]{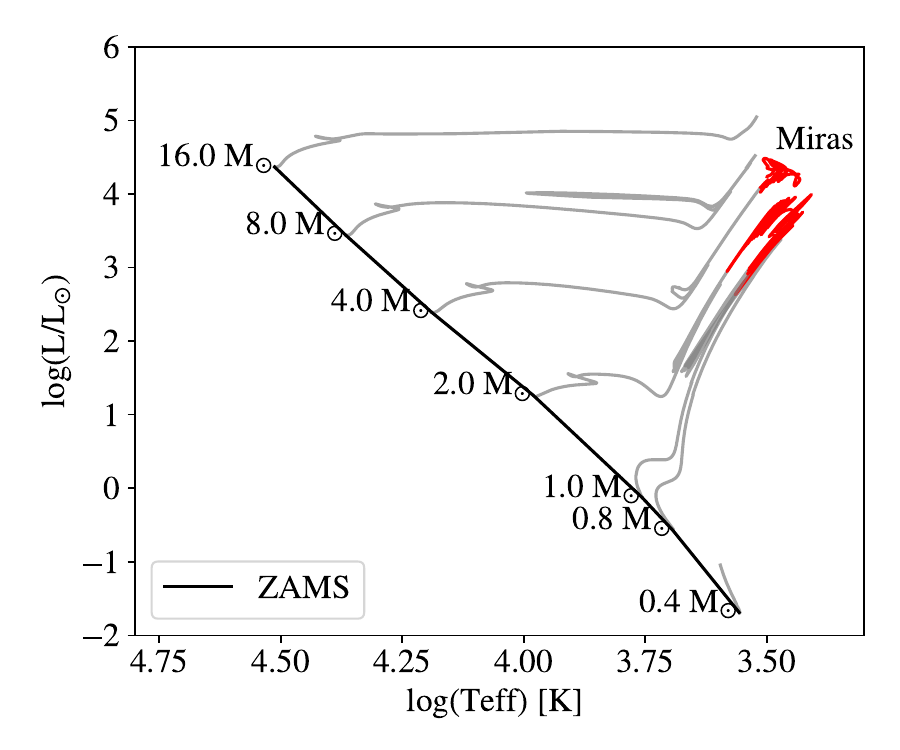}
%
%
\caption{Hertzsprung-Russell Diagram created using MIST evolutionary tracks \cite{Dotter_2016, Choi_2016, Paxton_2015}. The zero-age main sequence is shown in black, individual evolutionary tracks in gray, and the thermally-pulsating AGB evolutionary stage in red. Miras are located at the uppermost region of the thermally-pulsating AGB branch. 
}
\label{fig:HRDiagram}       
\end{figure}

\begin{figure}
\centering
\includegraphics[scale=0.1]{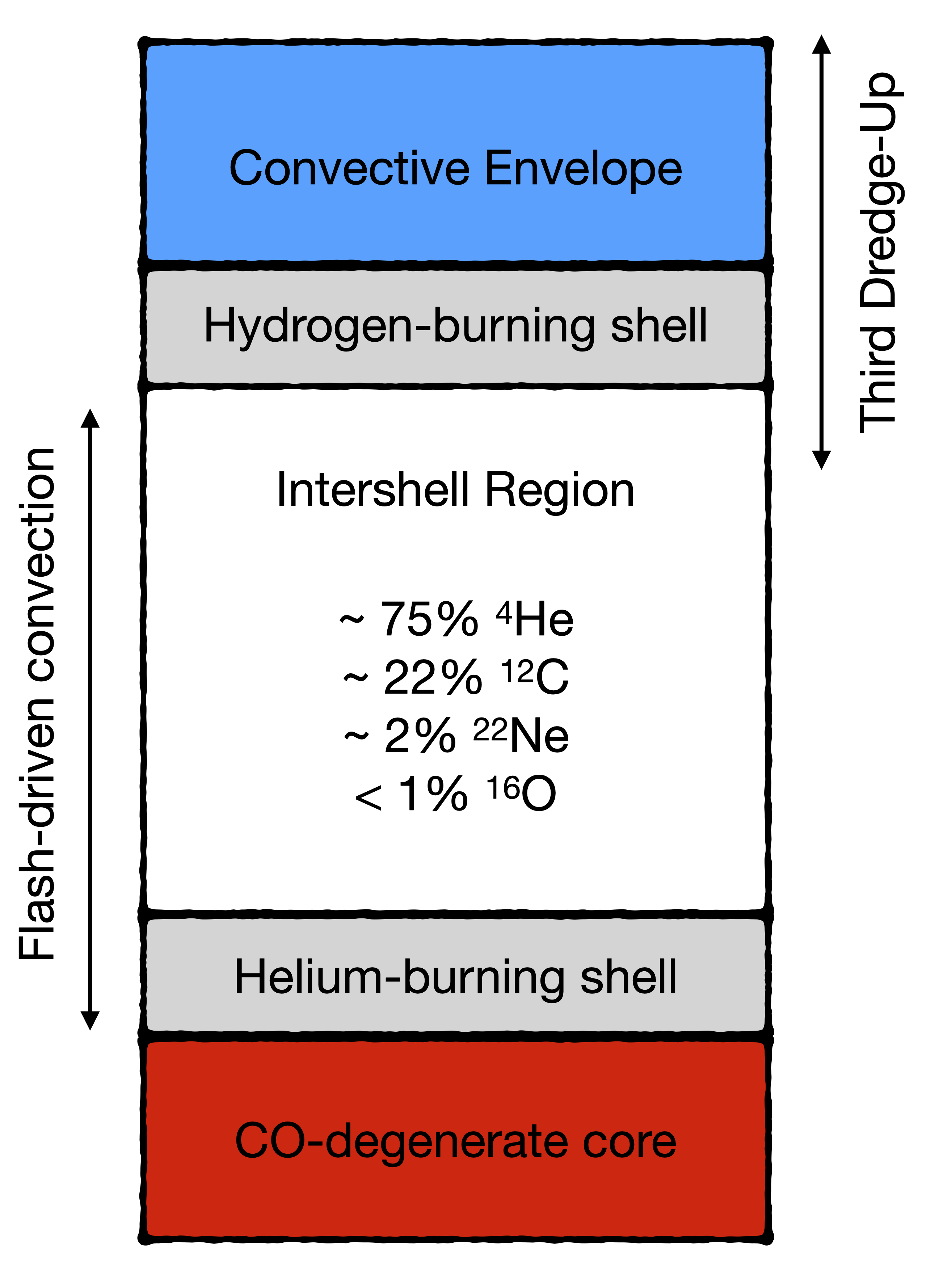}
%
%
\caption{The interior of an AGB star --- not to scale --- based on Figure 1 from \cite{Karakas_2002}. The degenerate CO core is surrounded by a He-burning shell. The He-burning shell is separated from the H-burning shell by an intershell region consisting mostly of helium and carbon. The H-burning shell lies directly below the deep, outer convective envelope. The ratio of radial thickness of the core compared to the envelope is $\sim 1 \times 10^{-5}$. 
}
\label{fig:TPcycle}       
\end{figure}

\section{Evolutionary Status}
\label{sec:evolution}

Miras are the brightest of the AGB stars and the AGB phase is the final stage of nuclear-burning life for vast majority of stars. It is often stated that stars with masses $0.8 M_\odot < M < 8 M_\odot$ will experience an AGB phase of evolution, though the exact upper limit depends on how convection is handled. The AGB stage is also brief, lasting $< 1\% $ of the main sequence lifetime. On the Hertzsprung-Russell diagram, these stars are typically identified as the bright stars above the tip of the Red Giant Branch (TRGB). Figure \ref{fig:HRDiagram} shows theoretical evolutionary tracks for stars from 0.4 $M_{\odot}$ to 16 $M_{\odot}$ and the location of the AGB, where the Miras reside. 

AGB stars have evolved past core hydrogen and helium fusion. Instead, they have a degenerate CO core and sustain themselves against gravitational collapse primarily through hydrogen-shell burning and quasi-periodic helium-shell flashes that ignite helium-shell burning. Stars actively undergoing this process are also known as thermally-pulsating (TP-AGB) stars. 

The cycle of thermal pulses is also responsible for dredging-up carbon from the core and potentially changing the surface chemistry of a star from O-rich to C-rich. AGB stars exhibit enrichments of elements produced through the slow neutron capture process (\textit{s}-process) and are also often experiencing rapid mass loss with rates of $ 10^{-8} M_{\odot} < \dot{M} < 10^{-4} M_{\odot}$ year$^{-1}$ \cite{Hofner_and_Olofsson_2018} and typical wind velocities of $\sim 10$ km/second. In evolved stars, mass loss and variability are also inextricably linked \cite{Marengo_2014}.

Stars ascending the AGB will are likely to experience multiple phases of pulsational instability, first as low-amplitude, overtone pulsators, and then finally as high-amplitude, fundamentally-pulsating Miras at the tip of the AGB. During the Mira phase, stars are at their most luminous and also have the highest mass-loss rates. This stage is also quite brief, lasting $\sim 10^5$ years for low-mass stars. Depending on their mass and abundance, due to the He-shell flashes, it is also possible that during their evolution, stars may also enter and leave the Mira phase more than once. 

The post-AGB evolution of a star is one of the least well-understood stages of stellar evolution. Low-to-intermediate mass stars are thought to become post-AGB stars and eventually blow off the outer layers to form planetary nebulae. The final life stage of more massive stars is even more uncertain, although they may go on to become super-AGB stars. 

There have been several excellent previous reviews of the AGB evolution, including \cite{Iben_1991, Frost_and_Lattanzio_1996b, Wood_1997, Busso_1999, Herwig_2005, Karakas_2014, Hofner_and_Olofsson_2018}. Here we will provide only a brief summary of the evolutionary status of Miras with a focus on the aspects most relevant to distance measurements. 

\subsection{The TP-AGB and Third Dredge Up}

A \textit{dredge-up} is the process that mixes the products of nucleosynthesis from the nuclear burning zone of the star to the its upper layers via convection. The first dredge-up occurs after core hydrogen burning but before helium burning starts, as the star is ascending the red giant branch. The second dredge up occurs for stars of 4-8 $M_{\odot}$ after core helium fusion comes to an end. Lower mass stars will not experience a second dredge-up at all. Therefore, the names of the dredge up processes do not necessarily indicate the order in which they occur. We will primarily focus on the \textit{third dredge-up}, which takes place when a star has entered the asymptotic giant branch, after a helium flash, and brings helium, carbon, and \textit{s}-process elements to the surface. 

During the AGB phase, the hydrogen-burning shell is the primary source of energy. As the star evolves up the AGB, the helium shell becomes thinner and thermally unstable, resulting in explosive, quasi-periodic ignitions of the helium-burning shell \cite{Schwarzschild_and_Harm_1965}. These are known as shell flashes or thermal pulses (TPs). Prior to the first TP, a star is in the early AGB phase and after the first TP, it is on the TP-AGB phase. 

Each of these cycles of AGB evolution can be broken down into four phases \cite{Iben_1981, Karakas_2014}:
\begin{enumerate}
\item{\textbf{Thermal pulse:} Helium-burning is ignited explosively at the base of the intershell region. High luminosities of $\geq 8L{_\odot}$ are produced for a short timescale of $\sim 10^2$ years. The energy from the flash creates a convective zone that extends from the He-burning shell to the H-burning shell and mixes the abundances in the intershell region.}
\item{\textbf{Power Down:} During power-down, energy produced from the thermal pulse expands the star, which cools the material outwards of the He-shell and halts fusion the H-shell. The surface luminosity will dip by $\sim 10\%$ during this expansion. The inner layers will cool, leading to a increase in stellar opacity.}
\item{\textbf{Third dredge-up:} The increase in stellar opacity results in the base of outer convective envelope moving inwards, past the previous H-burning shell and into the intershell region. This mixes the elements from He-burning to the surface, where they can be observed.}
\item{\textbf{Interpulse:} After the third dredge-up, the star will contract once again, reigniting the H-burning shell. The star then enters a long quiescent phase for $\sim 10^4$ years. During this time, the H-shell provides most of the surface luminosity before the cycle repeats.}
\end{enumerate}
Figure \ref{fig:TPcycle} shows the interior of an AGB star and the relative positions of the regions where these processes take place. 

Stars enter the AGB with a photospheric ratio of C/O $< 1$ (the exact ratio depends on the initial oxygen abundance) and are classified as O-rich (M-) stars. On the AGB, this can change through the third dredge up process, which brings carbon synthesized through the triple-alpha reactions at the base of the helium shell to the surface via a convective zone. If the surface C/O ratio becomes $> 1$, then the star is classified as C-rich (C-type) star. Environments with lower oxygen abundance are host to more C-rich stars, since less carbon is required to change the surface chemistry from O- to C-rich. 

The two subtypes can be readily separated in the basis of the molecular features in their spectra. Carbon and oxygen will form stable CO molecules, which then leaves only the element with greater abundance for other chemical reactions. 


\subsection{Hot-Bottom Burning}
\label{sec:HBB}

Another process that affects the surface chemistry of some AGB stars is hot-bottom burning. Hot-bottom burning occurs in relatively massive AGB stars $\sim 4-7 M_{\odot}$ and happens alongside the third dredge-up process. In stars experiencing hot-bottom burning, the surface convective envelope is deep enough to that proton-capture nucleosynthesis can occur at the base \cite{Bloecker_and_Schoenberner_1991, Lattanzio_1992, Boothroyd_1995, Marigo_2013}. 

When the temperature at the base of the convective envelope exceeds $5\times 10^7$K the CNO cycle is activated, which burns $^{12}$C to $^{14}$N. Furthermore, the entire convective envelope has a turnover time of 1 year, meaning that the entire envelope is well-mixed and will be exposed to this hot region within a year. This process prevents a C-rich atmosphere from forming and is responsible for limiting the number of C-rich variables at the longest periods and highest masses.

The exact mass at which this process activates is unknown \cite{Boothroyd_1992}. Models indicate that AGB stars with lower metallicity experience hot-bottom burning at lower initial mass compared to stars with higher metallicity. Therefore, the masses at which hot-bottom burning occurs -- and consequently, the masses of the stars that will become C-rich is not known and will almost certainly depend on the environment. 

\section{Period-Luminosity Relations}
\label{sec:PLRs}

The Period-Luminosity Relation itself is typically defined as a linear relationship between mean magnitudes of variables and the logarithm of their periods. The magnitudes can be determined through averaging the flux or magnitude mean or by the mean of a (generally sinusoidal for Miras) fit to fluxes or magnitudes. Fit means are typically more consistent than an average, and therefore, preferred. 

While mean magnitudes are by far the most commonly-used for P-L relations, Mira P-L relations also occasionally use bolometric magnitude or peak magnitude. Bolometric luminosities are more fundamental, easier to compare with models, and independent of circumstellar reddening, but they require more extensive observations over a longer period of time and in a large wavelength range. In addition, the results can be dependent on the color-dependent bolometric corrections used \cite{Whitelock_2009a}. Peak magnitudes, on the other hand, are sometimes used with optical observations of Miras \cite{Bhardwaj_2019, Ngeow_2023}. Due to the large amplitudes of these stars at shorter wavelengths, mean magnitudes can be difficult to measure and therefore peak magnitude can result in lower-dispersion P-L relations. In addition, the luminosity is \textit{inversely} dependent on period in the shorter optical wavelengths, and the P-L relation becomes roughly flat around the I-band. Figure \ref{fig:PLRs} from \cite{Ita_2011} shows the \textit{single-epoch} P-L relations for Large Magellanic Cloud Miras at a range of wavelengths, from optical to the far infrared. With mean magnitudes the scatter in the P-L relations would be significantly smaller. 

\subsection{Basics of Mira Pulsation}
\label{sec:pulsation}

Pulsating stars exist in various parts of the Hertzsprung-Russell Diagram, grouped in narrow (in color/temperature), nearly-vertical regions known as instability strips. While the relationship between luminosity and period for variable stars (specifically Cepheids) was determined empirically \cite{Leavitt_1912}, an intuitive explanation for the existence of the P-L relation can be derived by examining the relationship between pulsational period and radius. 

The boundary conditions for a pulsating star are similar to those in an organ pipe, where one end (the center) is closed, and the other (the surface) is open. In the fundamental, lowest-frequency mode, the entire star will contract and expand at the same time. Thus we expect that,
\begin{align}\label{eq:Persim}
{\rm\Pi} \sim 4 \frac{R}{\bar{v}_s} 
\end{align}
where ${\rm\Pi}$ is the period, $R$ is the radius, and $\bar{v}_s$ is the mean sound speed. Writing sound speed in terms of density gives us,
\begin{align}\label{eq:soundspeed}
    v_s = \sqrt{\frac{\gamma P}{\rho}}
\end{align}
where $\gamma$ is the specific heat ratio of the gas, $P$ is the pressure, and $\rho$ is the density. Then, taking the mean density $\bar{\rho} = M/(\frac{4}{3}{\rm\pi} R^3)$ (an approximation) and plugging this back into equation \ref{eq:Persim} we quickly find
\begin{align} \label{eqn:PipropRM}
{\rm\Pi} \propto R^{3/2}M^{-1/2}  
\end{align}
which indicates that pulsational period is strongly dependent on radius, but only weakly dependent on mass. 

To connect this more directly to the Period-Luminosity relation, we can rewrite $R$ in terms of $L$, the luminosity. We know that luminosity of a spherical blackbody is given by $L = 4{\rm\pi} R^2 \sigma T_{\rm eff}^4$. Assuming $T_{\rm eff}$ is fixed, then $L \propto R^2$. Now we can write equation \ref{eqn:PipropRM} as 
\begin{align}\label{eq:PipropLM}
   {\rm\Pi} \propto L^{3/4} M^{-1/2}
\end{align}
Luminosity is strongly dependent on mass, however mass is relatively insensitive to luminosity. Thus, we would find ${\rm\Pi} \propto L^x$, where $x \leq 3/4$. In practice, however, $T_{\rm eff}$ is not constant and thus $x$ can be $> 3/4$. Nonetheless, this indicates that there is a power law relationship between ${\rm\Pi}$ and $L$. As a result of this, P-L relations are typically shown and fit in log-log space, as a function of $\log$ period and magnitude. 

In order to compare theory with observations and determine the pulsational mode of a star, pulsation is sometimes described using the pulsational constant $Q$, which is defined as:
\begin{align}\label{eq:Q}
Q = {\rm\Pi} \left(\frac{M}{R^3}\right)^{1/2}
\end{align}
Here we solved for the constant term in equation \ref{eqn:PipropRM} which we previously ignored. In practice, we typically cannot measure $R$ directly, so we substitute $R$ with $L$ (luminosity) and $T_{\rm eff}$ (effective temperature) and get

\begin{align}
Q =  {\rm\Pi} \frac{M^{1/2} T_{\textrm{eff}}^3}{L^{3/4}} \times 5.13 \times 10^{-12} 
\end{align}
for $\Pi$ and $Q$ in days and $M$ and $L$ in solar units. However, $Q$ values can be difficult to calculate, because $L$ is the bolometric luminosity. For a Mira-like star with $T_{\textrm{eff}} \sim 3000$ K, $Q \sim 0.1$ days for fundamental pulsators and $Q\sim0.05$ days for overtone pulsators. In practice, some measured $Q$ values have been more than a factor of 2 off from the theoretical values. \cite{Wood_1990} argued that the differences in theoretical and measured $Q$ could potentially be attributed to metallicity (discussed further in Section \ref{sec:metallicity}). However, the magnitude of metallicity dependence needed to account for this difference would be quite significant: a difference of $0.4$ mag between a solar metallicity Mira one with metallicity comparable to the LMC. So far, such a strong dependence has not been observationally confirmed. 

This discrepancy likely lies in the pulsational models used. Recently, \cite{Trabucchi_2017, Trabucchi_2021} have shown that linear calculations of overtone long-period variable sequences match well with observations. However, nonlinear pulsational models of O-rich AGB stars are required in order to reproduce the large-amplitude fundamental mode pulsations of Miras. These nonlinear models provide significantly better agreement with observations, by reproducing the earlier onset of dominant mode pulsation and shorter periods at larger radii that are present in observed P-L relations. 



\subsection{Ages and Initial Masses of Miras}
\label{sec:ages}

Miras are the brightest stars in old or intermediate age populations. Similar to other fundamentally-pulsating variable stars, younger (and more massive) Miras are expected to have longer periods than older (and less massive) ones. This is known as the period-age (P-A) relation, and has been empirically motivated by stellar kinematics \cite{Feast_1963}. Theoretical investigations of the P-A relation for LPVs have generally been sparse due to the difficulty of modeling evolved red stars, but recent models have shown agreement with observations \cite{Trabucchi_and_Mowlavi_2022}. There is also evidence to suggest that C-rich Miras -- which are confined to a smaller period range than O-rich Miras -- may be slightly more massive than O-rich Miras at the same period and are generally younger stars. 

Observational evidence comes from a range of studies in the Galaxy and Local Group. O-rich Mira variables with log $P \sim 2.0 - 2.5$ days have been found in Galactic globular clusters \cite{Feast_2002}, which suggests that they have low initial mass. Comparisons of O-rich Mira velocity dispersion with the age and velocity dispersion relations of stars in the solar neighborhood indicates that stars with log $P \lesssim 2.3$ days have initial masses of $< 1 M_{\odot}$. Most Galactic O-rich Miras have log$P \sim 2.5$ days and age $\sim 7$ Gyr. More massive Miras with log $P \sim 2.65$ days are $\sim 3$ Gyr. 

C-Miras on the other hand, have mean log $P = 2.717$, age of $\sim 1.8$ Gyr, and initial mass of $1.8 \sim M_{\odot}$ \cite{Feast_2006}. \cite{Feast_2009} provides a more detailed analysis of the relationship between ages, periods, and kinematics.

\subsection{Variability Classification}
\label{sec:variability}

Miras are a part of a broader group of variables known as long-period variables (LPVs), which have periods ranging from $\sim 100- 1000$ days. Their P-L relations were first studied by \cite{Gerasimovic_1928}, nearly a hundred years ago. He compared the periods and mean visual magnitudes of 10 LPVs with trignometric parallaxes. LPVs are very common since nearly all stars on the red giant branch or AGB will exhibit some form of variability and many of these will fall under the classification of LPV. As noted by \cite{Soszynski_2013}, more luminous variables are also typically more variable, therefore the degree to which we can identify which stars are variable is dependent on the photometric precision of our measurements. 

LPVs can typically be grouped into one of three classes: Miras, semi-regular variables (SRVs), and slow irregular variables. As the names suggest, Miras are the most regular pulsators amongst LPVs while semi-regular variables and irregular variables often exhibit variations in their periodicity. Miras are typically distinguished from SRVs by their larger amplitudes. A visual-band peak-to-trough amplitude larger than 2.5 mag within a single pulsational cycle \textit{and} spectroscopic identification traditionally served as the cutoff between Miras and SRVs \cite{Kholopov_1985}. Recently, however, it has become more common to use only amplitude (often $\Delta I > 0.8$ mag, $\Delta K > 0.4$ mag) for classification since spectroscopic followup is limited and expensive \cite{Soszynski_2009}. While pulsational periods for Miras are generally very stable over many decades, with only an estimated 1\% of them showing variations \cite{Zijlstra_and_Bedding_2002}, it is more common to see stochastic variations in the cycle-to-cycle mean flux \cite{Mattei_1997, Whitelock_1997}. \cite{Huang_2018} estimated this to be $\sim 0.07$ mag in $H$-band.




\begin{figure}
\centering
\includegraphics[scale=0.7]{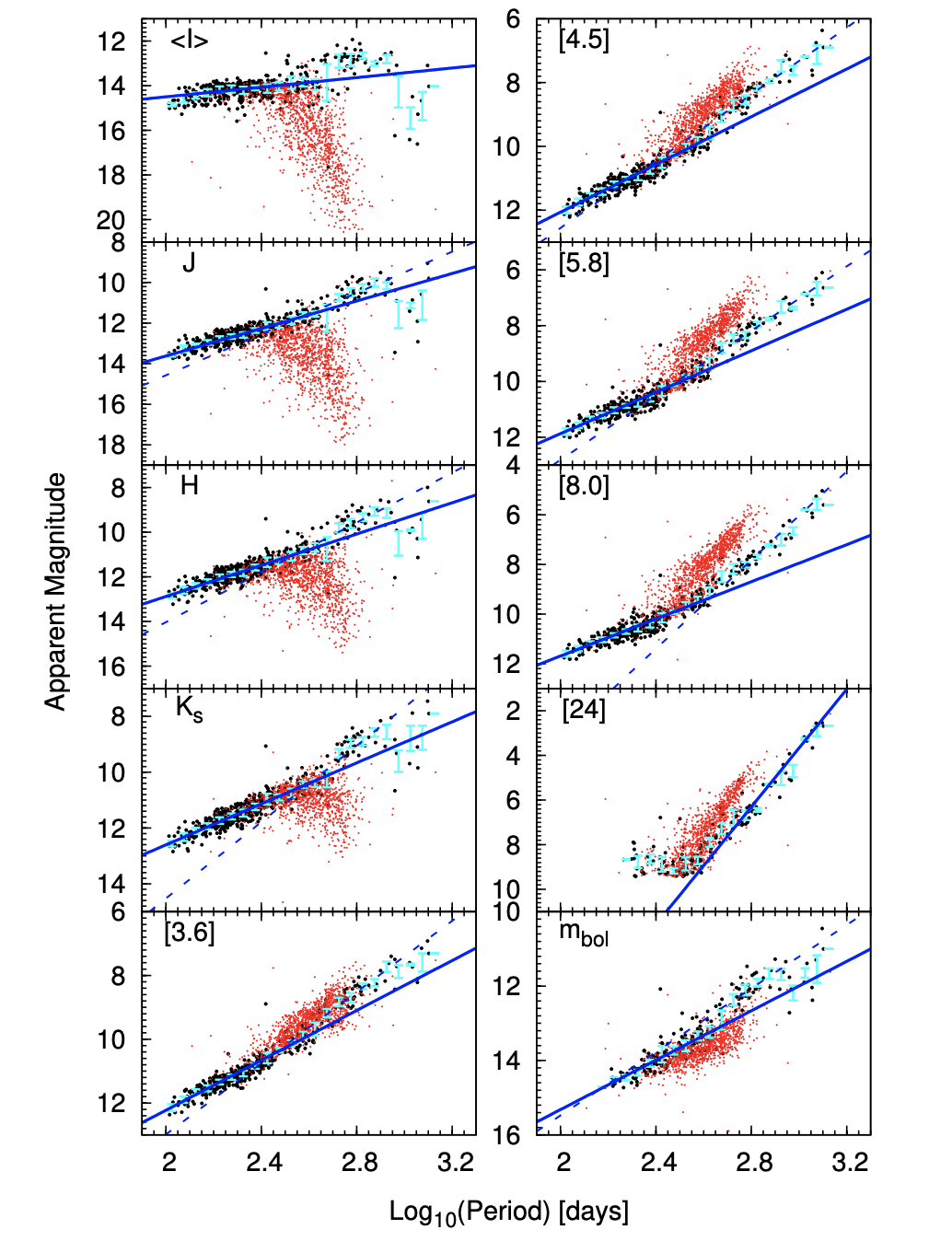}
%
%
\caption{Figure 2 from \cite{Ita_2011} showing the period-magnitude relations for OGLE III Miras in the LMC. The red points represent C-rich stars, while black points represent O-rich stars. The dashed and solid blue lines represent least-squares fits to the data. Mean magnitudes for the NIR and IR bands are derived from single-epoch observations, and thus show larger scatter.
}
\label{fig:PLRs}       
\end{figure}


\subsection{LMC}

The Mira variables in the Large Magellanic Cloud (LMC) are the best studied and have informed our knowledge of Miras in environments where the data is sparser. The LMC has been a critical galaxy both in the history and development of Mira P-L relations and also in the extragalactic distance ladder, as an \textit{anchor} galaxy (see Section \ref{sec:anchors}). 

With a small number of Miras from the Large Magellanic Cloud (LMC), \cite{Glass_Evans_1981} created the first P-L relation for Miras using bolometric magnitude. Subsequently, much of the history and study of Miras was a direct result of the availability of vast amounts of time-series photometry from microlensing surveys including the Optical Gravitational Lensing Experiment (OGLE, \cite{Udalski_2008}) and MACHO \cite{Alcock_1996}. Using MACHO data, \cite{Wood_1999} identified five parallel sequences (labeled $A-E$) on the Period-Luminosity plane, including Miras and SRVs. Miras were found to exist exclusively on sequence $C$ and were confirmed to be radial, fundamental mode pulsators, now in good agreement with theoretical models \cite{Trabucchi_2021}. 

Later, \cite{Glass_and_Lloyd_Evans_2003} determined that the $K$-band (2.2$\mu$m) Mira P-L relation exhibits a small scatter of $\sigma \sim 0.13$ mag. Using OGLE II and OGLE III time-series \cite{Udalski_2008} and 2MASS photometry, \cite{Soszynski_2005} found that sequence $C$ could further be separated into two -- $C$ and $C'$ in the period-Wesenheit index plane. With spectroscopically-confirmed variables, they were able to attribute this separation to O- and C-rich surface chemistries. \cite{Whitelock_2008} examined the $K$-band P-L relations for the O- and C-rich subtypes separately and found a similar level of dispersion ($\sigma = 0.14$ and 0.15 mag, respectively). This dispersion is also roughly comparable to the scatter in the Cepheid $K$-band P-L relation ($\sigma = 0.09$ mag, \cite{Macri_2015}). 

\cite{Ita_2011} showed that P-L relations for O-rich Miras in the LMC exist in a wide range of wavelengths, from $I$-band to [24] while C-rich Miras typically only exhibit a P-L relation at longer wavelengths (see Figure \ref{fig:PLRs}). Since then, numerous other studies have examined P-L relations in the from optical (\cite{Bhardwaj_2019}, using maximum light) to mid-infrared \cite{Iwanek_2021a}. $K$-band is typically the bandpass of choice for ground-based observations \cite{Whitelock_2012} thanks to the low intrinsic scatter of the P-L and the Mira spectral intensity peaking around $1-2\mu$m. There is also less circumstellar absorption in $K$ than at shorter wavelengths, and less circumstellar emission than at longer wavelengths. Thus, NIR P-L relations are the most widely studied, including \cite{Feast_1989, Wood_1999, Whitelock_2008, Ita_2011, Yuan_2017b}. 



\subsection{M33}\label{sec:M33}
M33 is another Local Group galaxy that has been extensively analyzed in variability studies and thus offers interesting comparisons between independent distance calibrators. There have been numerous Cepheid and TRGB distances to the galaxy, in addition to two JAGB distance determinations. \cite{Yuan_2017a}, which focused on the classification of the Miras in M33, the first to fit the Mira P-L using a quadratic relation and \cite{Yuan_2018} was the first to derive a distance to M33 using Miras with both linear and quadratic P-L relations. The quadratric form of the P-L was empirically motivated by the steeper slope and excess of above-PL stars at longer periods due to hot-bottom burning. A quadratic functional form also allows Miras of all periods to be fit simultaneously. However, this form has not been physically motivated. HBB models generally still predict a linear relation at long periods and previous studies had addressed this discontinuity by using two separate P-L relations. 

In \cite{Yuan_2017a}, Miras were classified using a Random Forest classifier and novel semi-parameteric periodgram method developed and described in detail in \cite{He_2016}. $I-$band light curves obtained from the M33 Synoptic Stellar Survey were combined with near and mid-infrared wavelength observations from the 3.8m UK InfraRed Telescope (UKIRT), the 4-m May-all telescope at Kitt Peak National Observatory (KPNO) and the 8-m Gemini North telescope. From this they obtained $\mu_{M33} = 24.80 \pm 0.06$ mag. 

Most recently, \cite{Ou_2023} used photometry from several surveys, including PANSTARRS, PTF, and ZTF, to create well-sampled optical light curves and derive a distance modulus of $24.67 \pm 0.06$ mag to the galaxy using O-rich Mira variables. They fit both maximum and mean light P-L relations. Miras were classified using their separation along the $J-K_s$ period-color plane. The majority (96\%) of periods were in agreement with those determined in the previous study by \cite{Yuan_2017a}.  

\subsection{NGC 5128}\label{sec:NGC5128}

NGC 5128 (Centaurus A) is the nearest giant elliptical galaxy and the first galaxy outside of the local group to have a Mira distance. It remains the only giant elliptical galaxy with a Mira distance. 

\cite{Rejkuba_2004} determined the distance to NGC 5128 using both Mira variables and infrared tip of the red giant branch (IR-TRGB) and found good agreement between the two methods, with $\mu_{Mira} = 27.92 \pm 0.19$, $\mu_{TRGBK} = 27.87 \pm 0.16$, and $\mu_{TRGBH} = 27.9 \pm 0.2$. Data used in the study taken using the ISAAC near-IR array on ESO VLT and absolute calibration for the Miras came from \textit{Hipparcos} parallaxes and the \textit{K}-band Mira P-L relation in the LMC. Observations consisted of 20 epochs of $K_s$-band observations spanning 1197 days and 1 epoch each of the $J_s$ and $H$ bands. Periods, amplitudes, and mean $K_s$ magnitudes were them calculated for the variables with 10 or more measurements. 

Only variables with regular periods, a sine-curve fit of $\chi^2 < 5$, and color bluer than $J_s - K_s \leq 1.4$ were used in the determination of distance. In addition, stars that were suspected to be aliased were not included in the fit. The mean log $P$ for the remaining sample was 2.5 ($\sim 300$ days).  The color cut removed LPVs with more circumstellar reddening and also matched the colors of the Miras used for the calibration of the Mira P-L relation in the LMC ($J-K < 1.5$ \cite{Feast_1989}). \cite{Rejkuba_2004} considered the possibility of hot-bottom burning in variables with $P > 400$ days. However, in practice, they also found that the distances determined with the full period range (up to just under 800 days at the longest end) and the shorter period variables only ($P < 400$ days) were nearly the same (differing by a few hundredths of a magnitude and well within the uncertainties). This could indicate that many stars, even at longer periods, are not actively undergoing hot-bottom burning. 

\subsection{Dwarf Galaxies}
\label{sec:dwarfs}

P-L relations also exist for some local group dwarf irregular galaxies, including NGC 6822 \cite{Whitelock_2013} and IC 1613 \cite{Menzies_2015}, which have distances determined using C-rich Miras. Perhaps most importantly for distance measurements, dwarf galaxies span a range of metallicities and star formation histories and thus are ideal environments for studying the relationship between stellar populations and galaxy evolution. While the C/O ratio is known to vary strongly with metallicity \cite{Battinelli_2005}, studies of the AGB P-L relations in dwarf galaxies with a wide range of metallicities have been instrumental in showing that there is no measurable metallicity dependence for the P-L relation \cite{Goldman_2019}. A more extensive review can be found in \cite{Whitelock_2018_review}.

\section{Measuring the Hubble Constant}\label{sec:H0}

Hubble's law states that: 
\begin{align}
v_r = H_0 \times D 
\end{align}
where $v_r$ is the recessional velocity and $D$ is distance. Hubble's Constant can be thought of as the constant of proportionality of $v_r$ and $D$. At large distances ($ z \gtrsim 1$), this relationship will begin to deviate from linearity depending on the cosmology of the universe. For nearby galaxies, local gravitational interactions can contribute significantly to or dominate the observed velocity. 

Galaxies used to measure H$_0$ must be within the ``Hubble flow". In other words, they must be in the regime where their recessional velocities are due almost entirely to the expansion of the Universe. In practice, this often means galaxies with distances $D > 100$ Mpc, or redshifts $z > 0.02$. Peculiar velocity corrections are typically applied to account for any remaining effects due to local gravitational interactions. 

However, geometric methods generally cannot reach galaxies in the Hubble flow. Even using Miras as secondary distance indicators, only the most luminous and massive stars would be able to reach these distances directly and their P-L relations are less well-understood and studied (see Section \ref{sec:longperiod}). A typical path towards measuring H$_0$ instead consists of a three-rung ladder. The first rung uses nearby (in the case of Miras, $D < 7.5$ Mpc) geometric anchors which determine the absolute magnitude of the Mira P-L relation. Short-period Miras (typically hundreds can be discovered per galaxy) are then used to reach $< 40$ Mpc distances, enough to measure the distances to and calibrate the peak luminosities of the nearest SN Ia, which are relatively rare (occurring at a rate of approximately 1 per galaxy per century). Thus, Miras serve as a bridge between the distances reachable by most geometric methods and the SN Ia in the Hubble flow that are capable of measuring H$_0$ directly. 

In the low-redshift ($z \sim 0$) limit, the intercept of the Hubble diagram $a_B$ is given by,
\begin{align}
a_B = \log cz - 0.2 m_B^0
\end{align}
where $m_B^0$ is the standardized (corrected for variations from fiducial color, luminosity, or host dependence) maximum-light apparent magnitude of the SNe. From this equation, we can see that $a_B$, which is determined solely from the third rung of the distance ladder, does not depend on any Mira distance measurements.

In order to solve for H$_0$, we combine the absolute SN Ia peak magnitude along with the intercept for the SN Ia Hubble diagram to write,
\begin{align}
\label{eq:H0}
\log H_0 = (M_B^0 + 5a_B + 25)/5
\end{align}
where $M_B^0$ is the fiducial SN Ia absolute magnitude. We determine $M_B^0$ by measuring the distance to local SN Ia host galaxies using Miras (or other primary distance indicators), and using the distance to convert the observed apparent SN Ia magnitudes ($m_B^0$) to absolute magnitudes ($M_B^0$). 



\subsection{Anchors}\label{sec:anchors}

The first rung of a cosmic distance ladder is often referred to as an \textit{anchor} because it sets absolute scale for the later rungs using geometry. The most obvious way to obtain a geometric distance is through parallax. However, in the case of highly-extended and evolved stars like Miras, the current consensus is that parallax measurements of AGB using \textit{Gaia} have large, asymmetric, and underestimated uncertainties and should be treated with caution (see Section \ref{sec:parallaxes} for a more in-depth discussion). Thus, we currently do not recommend using parallax measurements for calibrating the Mira H$_0$ measurement. 

Instead, since Miras are a ubiquitous, intermediate-age population, they can be discovered, and subsequently calibrated in almost any galaxy with a geometric distance. Mira-based H$_0$ measurements \cite{Huang_2020} rely primarily on geometric measurements to two anchor galaxies -- the LMC and NGC 4258 \cite{Huang_2018}. 



\subsubsection{NGC 4258}
\label{sec:NGC4258anchor}

NGC 4258 is a relatively nearby water megamaser host galaxy which has a geometric distance of $D = 7.576 \pm 0.082$ (stat.)$\pm 0.076$ (sys.) Mpc with 1.4\% precision \cite{Reid_2019}. Angular diameter distances to megamasers are obtained by combining spectral monitoring of the Keplerian motion of the masing clouds orbiting the black hole, along with VLBI mapping and a model of the maser disk geometry. Given that the galaxy is quite close, NGC 4258's peculiar motion is likely to contribute significantly to its recessional velocity. Thus, while it is unsuitable for direct H$_0$ measurements, it has served as an important anchor galaxy for Cepheids, TRGB, and Miras. 

\cite{Huang_2018} first calibrated the distance to this galaxy using a quadratic P-L relation based on \cite{Yuan_2017b} and the 2.6\% distance to the galaxy from \cite{Humphreys_2013, Riess_2016}. For P-L relation, 
\begin{align}
 m_{\rm F160W} = a_0 - 3.59(\log P - 2.3) - 3.40(\log P - 2.3)^2
\end{align}
where $m_{\rm F160W}$ is the magnitude in \textit{F160W}, $a_0$ is the zeropoint, and $P$ is period, in days.  \cite{Huang_2018} obtained $a_0 = -6.15 \pm 0.09$ mag as the absolute mean magnitude of a 200-day O-rich Mira using the quadratic relation. \cite{Huang_2020} later updated this  with the improved distance from \cite{Reid_2019} and used two linear P-L relations, 
\begin{align}
m_{\rm F160W} &= a_0 - 3.64(\log P - 2.3)
\label{eq:PLRH}\\
m_{\rm F160W} &= a_0 - 3.35(\log P - 2.3)
\label{eq:PLRF160W}
\end{align}
Equation \ref{eq:PLRH} used the linear slope derived for the $H-$band from \cite{Yuan_2017b} while Equation \ref{eq:PLRF160W} used a color transformation from $H$ to \textit{F160W} derived using O-rich Mira spectra to solve for a new slope. This resulted in $a_0 = -6.21 \pm 0.042$ mag for Equation \ref{eq:PLRH} and $a_0 = -6.25 \pm 0.042$ mag for Equation \ref{eq:PLRF160W}. 

The data for the Mira observations were taken using the HST WFC3/IR channel in two filters \textit{F125W} (HST wide $J$) and \textit{F160W} (HST wide $H$) in GO-13445 (PI: Bloom). Twelve epochs, spaced approximately monthly, were obtained of a single field over the course of a year. While the colors were originally intended to aid in separating the Miras into their O- and C-rich spectral subtypes, the color separation was found to be poor. Therefore, additional measures were taken to identify the two classes without using color information. Given that C-rich variables are rare at the shortest periods and also tend to have larger amplitudes, the sample was limited to $P < 300$ days and $0.4$ mag $< \Delta F160W < 0.8$ mag to reduce C-rich contamination. 

Of the 438 Mira candidates, 139 remained in the most stringent ``gold" sample. For the gold sample, optical data from a previous Cepheid observing campaign \cite{Macri_2006} was used to verify that the objects were variable in the \textit{F814W} (HST wide $I$) time-series. In addition, detection in \textit{F814W} was also taken as a proxy for color since the redder C-rich Miras were expected to be difficult to detect in the Cepheid data. However, despite these additional requirements, the zeropoints obtained for the full sample and gold sample were within 0.02 magnitudes, well within the uncertainties. 

Finally, the relative distance between NGC 4258 and the LMC was calculated after applying a Mira-specific color correction to ground-based $H$ band P-L relation. This was found to be in agreement with the geoemtric distances for the two galaxies. 

\subsubsection{LMC (again)}
\label{sec:LMCanchor}

\cite{Huang_2020} used the 1.2\%  \cite{Pietrzynski_2019} distance to the LMC of $\mu_{\rm LMC} = 18.477 \pm 0.004 {\rm (stat)} \pm {\rm (sys)}$  mag to calculate an absolute calibration of $a_0 = -6.27 \pm 0.03$ mag for Equation \ref{eq:PLRF160W}, consistent to within the uncertainties with the zeropoint of the Mira PLR from NGC 4258. 

While LMC Miras were not directly observed with HST WFC3/IR, using the OGLE-III sample of O-rich Mira variables \cite{Soszynski_2009}, \cite{Huang_2018} determined the relative distance modulus between the LMC and NGC 4258, 
\begin{align}
\mu_{\rm N4258} - \mu_{\rm LMC} = 10.95 \pm 0.01 {\rm (stat) } \pm 0.06 {\rm (sys)\  mag.}
\end{align} 
which is consistent with the relative Cepheid distance between these two galaxies. 

In order to directly compare the two samples without observations in the same bandpasses, \cite{Huang_2018} derived an $H$ to $F160W$ transformation of 
\begin{align}
F160W = H  + 0.39(J-H)
\end{align}
using O-rich Mira spectra from \cite{Lancon_and_Wood_2000} as input into PySynPhot \cite{STScI_2013}. 

In addition, the sample of Mira variables from OGLE-III also included variables with a wider range of amplitudes and periods than those used in the NGC 4258  sample. Thus, the OGLE-III sample was restricted to only O-rich Miras with $ 240 < P < 400 $ days and $0.4 < \Delta H < 0.8$ mag to provide a closer match to the HST sample for the calibration.  

\subsection{SN Ia Calibrators}
\label{sec:SNIaCalibrators}

The second rung and largest contributer to the overall H$_0$ error budget is the calibration of the SN Ia peak luminosity with Miras in local SN Ia host galaxies. To date, only two SN Ia have had luminosities calibrated with Mira distances. 

\subsubsection{NGC 1559}
\label{sec:NGC1559}

\cite{Huang_2020} determined the distance modulus to NGC 1559, the host of SN 2005df to be $\mu_{N1559} = 31.41\pm 0.050 {\rm (stat)} \pm 0.060 {\rm (sys)}$ using a sample of 115 Miras. At $19.1 \pm 1.1$ Mpc, NGC 1559 is the most distant galaxy in which Miras have been studied. The distance for this galaxy was used by \cite{Huang_2020} to determine a fiducial SN Ia luminosity of $M_B^0 = -19.27\pm 0.13$ mag, and in agreement with the \cite{Riess_2022} Cepheid measurement of $M_B^0 = -19.360 \pm 0.106$ mag for this galaxy. Using both the NGC 4258 and LMC P-L relations as anchors, this results in H$_0 = 73.3 \pm 4.0$ km s$^{-1}$ Mpc $^{-1}$. The methodology laid out in this paper is the blueprint for measuring Miras in other local SN Ia hosts.  However, because this H$_0$ measurement uses only one SN Ia calibrator, the 5\% uncertainty is dominated by the statistical uncertainty in the peak magnitude of the single SN Ia and the result is within 2-$\sigma$ agreement with both \textit{Planck} and the most precise Cepheid measurement.


Data used for the NGC 1559 study consisted of 10 epochs of WFC3/IR channel \textit{F160W} observations (GO-15145; PI: Riess) with roughly monthly spacing and a baseline slightly longer than 1 year. Unlike most Mira studies, the NGC 1559 Mira sample did not use color as a selection criterion since \cite{Dalcanton_2012, Huang_2018} showed that HST wide-band colors are an ineffective discriminant of both O- and C-rich AGB stars in general and Miras in particular. Instead, amplitude and period cuts were used to limit the C-rich contamination. The remaining C-rich contamination (estimated to have an bias of $\sim -0.07$ mag in the worst-case) was estimated by using the LMC Mira sample to model a pure O-rich sample and a C-rich contaminated sample. Because C-rich Miras are expected to be fainter in \textit{F160W} than O-rich Miras and have a different period distribution than O-rich Miras, the shape of the zeropoint-vs-period curve ($Z(P)$) is different for a pure O-rich sample and a sample contaminated with C-rich Miras. The contaminated sample is then scaled to fit the $Z(P)$ relation of NGC 1559, yielding a correction of $-0.057 \pm 0.024$ mag and no correction when applied to NGC 4258. This method was tested on the M33 Mira sample from \cite{Yuan_2018} by creating a contaminated population and then correcting the zeropoint. The correction of $-0.04 \pm 0.01$ mag was found to be in agreement with the true value of $-0.034$ mag. 


\subsubsection{M101}
\label{sec:M101}

M101 is the most recent SN Ia host galaxy to have a distance measurement using Miras \cite{Huang_2023}. 288 O-rich Mira candidates were discovered in the field of SN 2011fe using a long baseline of HST observations, spanning $\sim 2900$ days. Many of the observations were originally taken to the study the late-time lightcurve of the SN, but 5 follow-up observations were obtained to improve the spacing and recovery of the shorter-period Miras. From the sample of 211 Miras with periods between 260 and 400 days, the distance modulus to M101, using the LMC and NGC 4258 as anchors, was determined to be $\mu_{\rm M101} = 29.10 \pm 0.06 $ mag. In combination with the SN Ia calibration from NGC 1559, this yielded $M_B^0 = -19.27 \pm 0.09$ mag. The resulting H$_0$ measurement combining both SN Ia calibrators is $72.37\pm 2.97$ km s$^{-1}$ Mpc$^{-1}$, and confirmed previous indications that the local universe value of $H_0$ is higher than the early-universe value at $\sim 95\%$ confidence. This 4\% measurement is still dominated by the statistical uncertainty in the peak luminosity of the two SN Ia calibrators. 

As the nearest SN Ia host galaxy with modern (CCD) photometry, the distance to this galaxy has also been measured numerous times using Cepheids and TRGB, offering a opportunity for comparison with of the Mira distane with the two most commonly-used primary distance indicators. The recent Mira measurement falls within 1-$\sigma$ of the mean of the recent Cepheid and TRGB measurements as well as and within $1-2\sigma$ of all of the individual  Cepheids and TRGB measurements. 


\subsection{Methodology}
\label{sec:Methodology}

The methodology we describe here applies primarily to the targeted Mira searches in galaxies calibrated using sparse HST data. Ground-based observations have used diverse observing strategies and typically prefer $K$-band for P-L relations. However, Mira H$_0$ measurements have necessitated the use of HST to resolve stellar populations in SN Ia host galaxies and therefore use a slightly shorter bandpass -- WCF3/IR's \textit{F160W} ($\sim 1.6\mu$m). This is the longest wavelength wide-band filter available with HST and also coincides well with the peak of a typical Mira's spectral energy distribution.

\subsubsection{Observations}
\label{sec:Photometry}

Observational baselines for Miras span around one year or slightly longer and typically consist of $\sim 10$ epochs, with roughly monthly spacing. To reduce aliasing, it is generally preferable to space the observations such that the Miras at the target period ranges (usually $\sim100-400$ days) will be sampled a minimum of three phase points in one cycle. 

The initial steps of the variability search are similar to the process for Cepheids described in several SH0ES papers including \cite{Hoffmann_2016, Yuan_2021, Yuan_2022}. Once the data is collected, a variability search is performed to cull as many nonvariable objects as possible while retaining the variables. Crowded-field photometry is used to create lightcurves for the candidates, which are then cut down to a sample of O-rich Mira candidate variables using selection criteria. 

The major difference in the analysis process between Cepheids and Miras lies in the wavelengths used for determining the positions for the variables. Miras are very red stars and therefore typically not observed in optical wavelengths. With Cepheids, the finer-resolution optical observations are generally used to classify the stars and pinpoint their locations. However, given that they are an older population, Miras are generally found in less dusty areas than the younger Cepheids, and thus this entire search is performed only in the NIR. 

\subsubsection{Selection Criteria}
\label{sec:SelectionCriteria}

Selection criteria are used to separate the Miras from SRVs and other potential contaminants. Unlike Cepheids, which have a distinctive sawtooth shape in optical wavelengths and are often identified on the basis of their fit to templates, there currently are no templates for Miras and their light curve shapes of Miras in optical wavelengths. However, even in the optical, most ($\sim 80\%$) short-period Miras  do not show substantial deviations from a symmetric light curve \cite{Vardya_1988}, and $\sim$70\% of them do not significantly deviate from a purely sinusoidal shape \cite{Lebzelter_2011}. 

\begin{figure}
\centering
\includegraphics[scale=0.40]{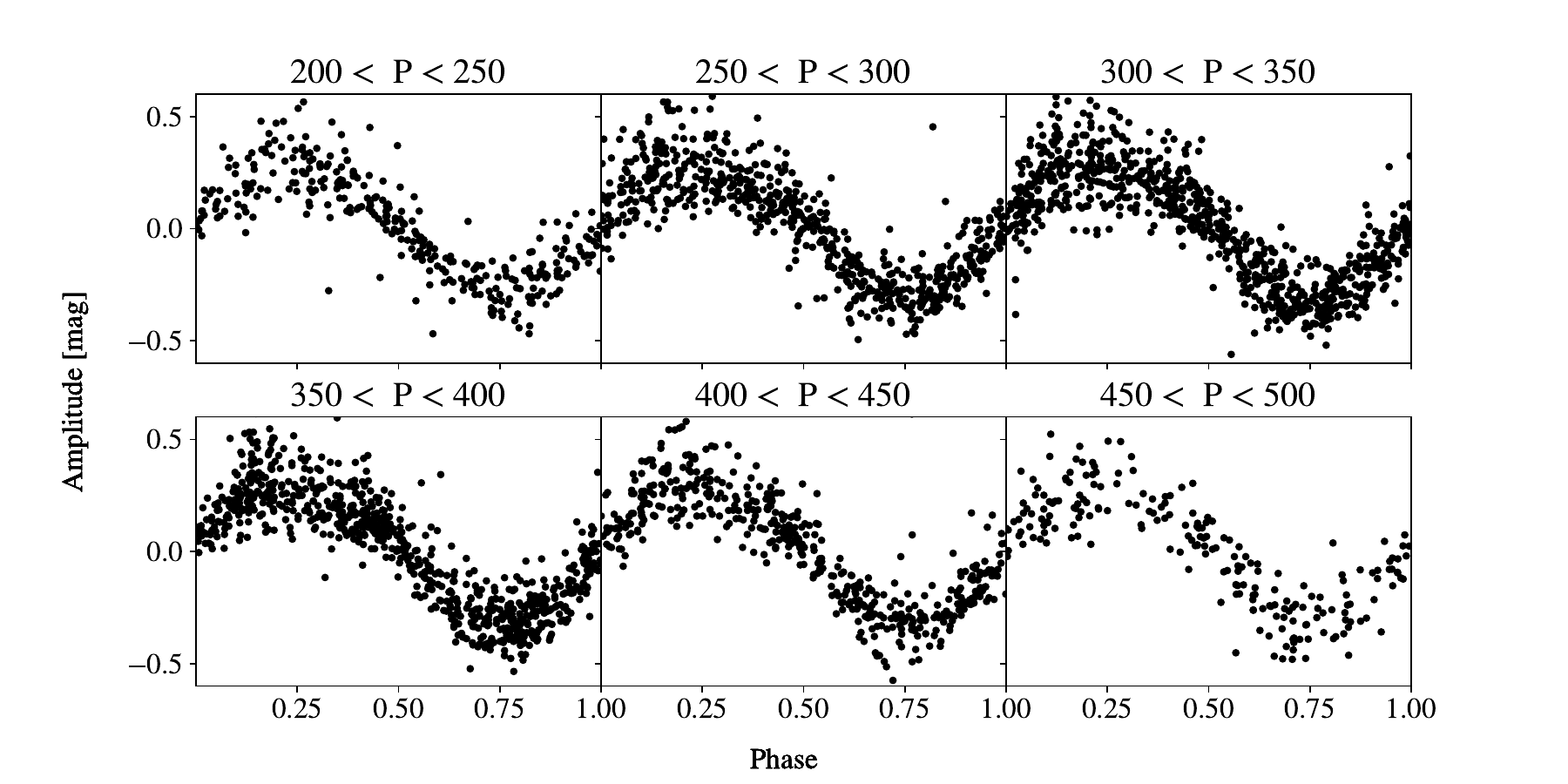}
%
%
\caption{Figure 3 from \cite{Huang_2023} showing the stacked lightcurves of $\sim 300$ Miras in the SN Ia host galaxy M101. Individual observations are shown with their fit phase, mean magnitude is subtracted, but amplitudes are not scaled.
}
\label{fig:lightcurves}       
\end{figure}

Similarly to Cepheids, the lightcurves of Miras in NIR are even more sinusoidal (see Figure \ref{fig:lightcurves} showing composite light curves at a range of periods). Therefore, the period, amplitude, and mean magnitude of Miras are typically fit using only a sine function. When \cite{Huang_2018} applied a Bayesian information criteria before adding additional harmonics, they found that $\sim 99\% $ of Miras favored a first order fit. 

Amplitude cuts are used to remove SRVs or potential C-rich Miras (which often have larger amplitudes). An initial period cut is also typically applied to remove objects with periods that are either too short ($< 100$ days) or more likely to be C-rich or hot-bottom burning ($>300-400$ days). A summary and comparison of the cuts used in a few recent studies is shown in Table \ref{tab:selection_criteria}. 

At this stage, the remaining objects are likely to be Miras, but further corrections may be applied to remove objects that are found to be particularly crowded (either through surface brightness cuts or cuts on the crowding corrections) or below the completeness limit and thus may bias the zeropoint of P-L relation. C-rich contamination corrections are also typically applied, with \cite{Huang_2020} estimating the level to be a few hundredths of a magnitude.





\begin{table}
\caption{Mira Selection Criteria}
\label{tab:selection_criteria}       
\begin{tabular}{p{2cm}p{3cm}p{3cm}p{3cm}}
\hline\noalign{\smallskip}
 & M101 & NGC 1559 & NGC 4258 (gold)  \\
\noalign{\smallskip}\svhline\noalign{\smallskip}
Period Cut &  200 d $< P <$ 500 d  & $240$ d $< P < 400$ d &  $P < 300$ d \\
Amplitude Cut & $0.4 \text{ mag} < \Delta \emph{F160W} < 0.8$ mag & $0.4 \text{ mag} < \Delta \emph{F160W} < 0.8$ mag & $0.4 \text{ mag} < \Delta \emph{F160W} < 0.8$ mag \\
Color Cut & $m_{F110W} - m_{F160W} < 1.3$ mag & --- & $m_{F125W} - m_{F160W} < 1.3$ mag \\
Surface Brightness Cut & ---  & \surfbrightness counts/second &  --- \\
$F$-statistic : & $\chi^2_{\text{s}}/\chi^2_{\text{l}} < 0.5$ & $\chi^2_{\text{s}}/\chi^2_{\text{l}} < 0.5$ & --- \\
\textit{F814W} Detection & --- & --- & Slope-fit to \textit{F814W} data $> 3\sigma$\\
\textit{F814W} Amplitude & --- & --- & $\Delta$ \textit{F814W} $> 0.3$ mag\\
\noalign{\smallskip}\hline\noalign{\smallskip}
\end{tabular}

A comparison of the criteria for the final Mira sample in M101 with those used for NGC 1559 and NGC 4258. Owing to differences in available data, we were unable to match the criteria exactly between all of the observations.
\end{table}


\subsection{Other Considerations}

\subsubsection{Metallicity}
\label{sec:metallicity}

As touched upon in Section \ref{sec:pulsation}, theory suggests that the Mira P-L relation may have a dependence on metallicity. \cite{Wood_1990} noted that if two stars with the same mass but different metallicities begin to pulsate at the same radius (and thus, period), they will exhibit different luminosities. However, this assumes that these stars would become unstable to Mira-like pulsations at the same radius. It is also possible that Miras with the same mass but different metallicity would begin to pulsate at different radii, which may have the effect of mitigating the change in luminosity. Thus far, the agreement between theory and observations of Miras in a wide range of metallicities suggests that the dependence of the zeropoint of the P-L on metallicity is weak. For a more detailed explanation, see \cite{Feast_1996}.  

As discussed in Section \ref{sec:dwarfs}, dwarf galaxies are the ideal testing grounds to understand the connection between galaxy evolution and stellar populations. Most recently, \cite{Goldman_2019} examined the IR P-L relations of AGB stars in environments with metallicity ranging from 1.4\% solar metallicity  ([Fe/H] = -1.85) to 50\% solar metallicity ([Fe/H] = -0.38) and found no discernible difference in the pulsational properties of stars in metal-poor and metal-rich environments. Similarly, \cite{Bhardwaj_2019} also found that the mean $K_s$ band O-rich Mira measurement between the LMC and SMC showed no sign of a metallicity effect. 

Other works have put tentative upper limits on the metallicity effect by showing that agreement in Mira P-L relations without metallicity corrections is within a few hundredths of a magnitude in the NIR. A study of the $K$-band P-L relation of Miras in the LMC, Galactic globular clusters, and Galaxy \cite{Whitelock_1994} showed agreement to within the measurement uncertainties. \cite{Whitelock_2008} found an upper limit of $\sim 0.1$ mag in the zeropoint of LMC and Galactic O-rich Miras with \textit{Hipparcos}  and VLBI parallaxes and those in globular clusters. However, when \cite{Huang_2018} used the same data, but an updated LMC distance from \cite{Pietrzynski_2019} this discrepancy dropped to 0.01 mag. For HST \textit{F160W}, the wide-band filter used for H$_0$ measurements, \cite{Huang_2018} found $\sim0.03$ mag agreement (within errors) in the relative distance modulus between the LMC and NGC 4258 determined using Miras that from Cepheids and from geometry.  Together, this suggests that the AGB P-L relations are a reliable tool for determine distances regardless of metallicity, at least down to the $\sim0.03$ mag level in the NIR. 

\subsubsection{Parallaxes}
\label{sec:parallaxes}

While other distance indicators such as Cepheids or RR Lyrae have been calibrated using \textit{Gaia} parallaxes, parallax measurements are difficult to obtain for large, red, evolved stars such as Miras using an optical telescope like \textit{Gaia}. Typical radii for Miras range from $250 R_{\odot} - 800 R_{\odot}$ \cite{van_Belle_2002}. \cite{Xu_2019} found that the redder the star, the larger parallax uncertainties in \textit{Gaia} DR2. Red stars, including AGB stars, typically have more complex surface dynamics, larger physical size, and more dust. Empirically, the Galactic Mira PLR using \textit{Gaia} shows large scatter and shows possible disagreement with previous calibrations \cite{Sanders_2023}. 

\textit{Gaia} parallax uncertainties for Miras and AGB stars in general are both underestimated (up to a factor of 5.44 for sources with $G < 8$ mag) and asymmetrical. \cite{Andriantsaralaza_2022}  caution against using parallaxes when uncertainties are larger than 20\% and employing the \textit{Gaia} catalogue parameter RUWE (re-normalised unit weight error) as the sole measure of quality of \textit{Gaia} astrometric data for AGB stars. They were able to obtain distance measurements to AGB stars using maser observations with very long baseline interferometry (VLBI). Unfortunately, this method can only measure distances to Miras with substantial circumstellar envelopes, and is typically not applicable to the Miras currently used for the distance ladder. However, it is appropriate as a cross-check to the AGB distances obtained by \textit{Gaia}.

Difficulties in parallax measurements with Miras were not unexpected and several factors -- both astrophysical and instrumental, contribute. Physically, Mira radii are larger than 1 AU -- in other words, their angular diameters are larger than their parallaxes. Both angular size and parallax scale as $1/d$, where $d$ is distance to the star. Thus, the physical size of a star is a problem whether the Mira is located nearby or far away. On its face, this alone does not appear to be particularly problematic for parallax measurements since an astrometric solution can be found using only on the motion of the photocenter of the extended source. However, Miras are not uniformly-illuminated discs. Moreover, they have complex stellar surface dynamics caused by large convective cells that can result in the motion of their stellar photocenters \cite{Chiavassa_2018}. Bright sources, including Galactic Miras, can also reach instrumental saturation which results in less accurate astrometry \cite{ElBadry_2021}. 

\textit{Hipparcos} had more success obtaining parallaxes for Miras. \cite{Whitelock_2000b} examined subsets of the data and found that the best solution excluded short-period red group stars and also low-amplitude variables. \cite{Whitelock_2008} used the revised \textit{Hipparcos} parallaxes from \cite{vanLeeuwen_2007} data to obtain a calibration for Galactic, O-rich Mira variables in the $K$ band of $\delta = -7.25 \pm 0.07$ mag.


\section{Long-Period Miras}
\label{sec:longperiod}

The long-period end of the Mira distribution is less studied than the short-period objects discussed throughout the this review for a number of reasons. In addition to more complicated physics (including the effects of hot-bottom burning, heavier mass-loss, and larger circumstellar envelopes compared to short-period Miras), they require extended observational baselines to accurately determine their periods and are formed from stars of higher initial mass and are therefore considerably rarer, younger, and more likely to be located closer to dusty regions where they were born than short-period Miras.  Additionally, OH/IR stars, which are bright infrared sources with Type II OH masing emission and intermediate initial mass, fall into this category \cite{Engels_1983} and some of these extreme objects have periods as long as 1800 days. Long-period Miras also often follow a different P-L relation than their shorter-period counterparts.

Despite these difficulties, there remains great interest in studying long-period Miras both for distance ladder applications and as tracers of structure \cite{Urago_2020}. Assuming the same P-L slope as the shorter-end, their $F160W$ luminosities will \textit{start} at $-7.3$ mag on the low ($400$ days) end. As seen in Figure \ref{fig:PLRs}, the slope for these longer-period objects can often be steeper than at the short-period end, likely due to hot-bottom burning \cite{Whitelock_2003}. Thus, the brightest objects can be up to 2.5 magnitudes more luminous and reach distances $>50$ Mpc with Hubble Space Telescope. Long-period objects also appear to have tighter P-L relations in the mid-infrared which may be particularly advantageous with James Webb Space Telescope observations. 

Very long period ($P\gtrsim 1000$ day) Miras have been discovered locally in the SMC, LMC, and Galactic Bulge by OGLE \cite{Soszynski_2009, Soszynski_2011, Soszynski_2013} and in the nearby galaxies Sgr dIC and NGC 3109 \cite{Whitelock_2018, Menzies_2019} using the $JHK_s$ bands from the InfraRed Survey Facility (IRSF). The SPIRITS \cite{Kasliwal_2017, Karambelkar_2019} and DUSTiNGS \cite{Boyer_2015a, Boyer_2015b, Goldman_2019} collaborations have also found hundreds of long-period variables using 3.6$\mu$m and 4.5$\mu$m that are likely Miras in a wide range of host environments using \textit{Spitzer} Space Telescope some of them with $P\sim 1000$ days. 


\section{Comparison with other Indicators}
\label{sec:comparison}

\begin{table}
\caption{Comparison of Distance Indicators}
\label{tab:comparison}       
\begin{tabular}{p{2cm}p{2.5cm}p{2.5cm}p{2.5cm}p{2.5cm}}
\hline\noalign{\smallskip}
\textbf{Indicator} & \textbf{Age} & \textbf{Time-Series} & \textbf{mag$_{H}$}  & \textbf{Optical Data}\\
\noalign{\smallskip}\svhline\noalign{\smallskip}
\textbf{Miras:} & $2-8$ Gyr & Yes$^a$ & $< -6.25^b$ & No\\
\textbf{Cepheids:} & $ \lesssim 100$ Myr & Yes$^c$ & $< -6.1^d$ & Yes\\
\textbf{TRGB:} & $\gtrsim 4-5$ Gyr  & No &  --- & Yes \\
\textbf{IR-TRGB:} & $\gtrsim 4-5$ Gyr  & No &  [$-5.5, -6.5]^e$ & No \\
\textbf{JAGB:} & 300 Myr $- 1$ Gyr & No & $ -6.2$ [$J$], $-6.8$ [$H]^f$ & No\\
\noalign{\smallskip}\hline\noalign{\smallskip}
\end{tabular}

A comparison of the stellar distance indicators that have been shown to be capable of reaching SN Ia host galaxy distances. For the sake of brevity, Miras refers to O-rich, short period ($P \lesssim 400$ days) Miras. 
\\
$^a$ timescale $> 1$ year; $^b$ for $P > 200$ d; $^c$ timescale $> 60$ days; $^d$ for $P > 15$ d; $^e$ IR-TRGB luminosity depends strongly on metallicity; $^f$JAGB is most-commonly studied in the \textit{$J-$}band (calibration from \cite{Madore_2020}) but a recent work by \cite{Li_2024} provided a number of $H$-band absolute magnitude calibrations.
\end{table}

While there are numerous excellent stellar distance indicators, we will limit this comparison to those that have currently been shown to be capable of bridging the gap between nearby, geometric measures and the local host galaxies of SN Ia. This includes Cepheids, the Tip of the Red Giant Branch (TRGB), J-Region Asymptotic Giant Branch (JAGB) stars \cite{Lee_2021}, and short-period Miras and typically excludes fainter stellar indicators such as RR Lyrae and Type II Cepheids. For brevity, ``Miras" refers exclusively to short-period O-rich Miras. 





\noindent\textbf{Luminosity:} Cepheids and short-period Miras are roughly comparable in luminosity in the NIR. Both are $\gtrsim 10^4 L_{\odot}$ in \textit{H}-band. JAGB (which includes C-rich Miras) are brighter than the shortest-period Miras ($P < 200$ days), but typically fainter than the brightest Miras. As a result, despite requiring time-series photometry, Miras and Cepheids can be more efficient than TRGB and JAGB larger distances (e.g. $D > 10-15$ Mpc). It is necessary to resolve stars $0.5 - 1.0$ mag below the tip/main clump for TRGB and JAGB measurements. On the other hand, peak-to-trough amplitudes of Cepheids and Miras generally span $< 0.3$ mag and $0.4 - 0.8$ mag respectively, and must also be detected in order to determine their mean magnitudes.


\noindent\textbf{Age:} Cepheids are young, massive stars, typically found clustered near dusty spiral arms and star-forming regions and are on average more crowded than the older indicators. They are generally found in the disk of spiral galaxies. JAGB are C-rich stars and trend younger than O-rich Miras, but are still found in many galaxy morphologies. TRGB are a ubiquitious older population and are best observed in the halos of galaxies \cite{Beaton_2018, McQuinn_2019}. Miras are also a ubiquitous older population. 


\noindent\textbf{Time-Series Requirements:} A major advantage that TRGB and JAGB have over Cepheids or Miras is that they do not require time-series observations since both are identified using their positions on a color-magnitude diagram (CMD). This makes TRGB and JAGB more efficient for nearby galaxies $ D \lesssim 10-15$ Mpc, since only a single epoch of observation is necessary. 


\section{Conclusions and Future Prospects}\label{sec:conclusions}


As the most luminous of the intermediate-to-old stellar distance indicators, Miras are in a unique position to expand the sample of local SN Ia with calibrated peak magnitudes and are also the only distance indicator capable of cross-checking all of the others. They will be particularly exciting in the upcoming decade with the increase focus on NIR and IR observations. 

Presently, due to the small number of SN Ia calibrators, the uncertainty in the Mira-based H$_0$ is dominated by the statistical uncertainty in the peak magnitude of SN Ia. The best path forward to improving H$_0$ measurements with Miras is to increase the number of SN Ia that have been calibrated with Miras. This can be done either by revisiting the local host galaxies that already have Cepheid distances -- allowing for a direct comparison of Mira and Cepheid distances -- and by searching for them in the galaxies of SN Ia that have not yet been calibrated with other indicators due to their distance, or lack of star formation. 

James Webb Space Telescope (JWST), while not optimally suited for obtaining time-series, is able to photometrically separate O- and C-rich Miras in SN Ia host galaxies with medium-band filters and will also be able to un-crowd the backgrounds of previously-observed variables, similar to what has currently been done for Cepheids \cite{Riess_2023}. JWST also has the sensitivity to potentially measure a two-rung H$_0$ using geometry and Miras only -- without the inclusion of SNe Ia. However, it suffers from large overheads that make multi-epoch visits undesirable. 

The upcoming Vera C. Rubin Observatory's Legacy Survey of Space and Time will cover nearly half of the sky in $u, g, r, i, z, $ and $y$ filters. Rubin has the potential to be particularly useful in developing the first rung of the Mira distance ladder, since it is expected to discover variables in galaxies out $< 15$ Mpc \cite{Yuan_PhD}. While this is too small of a volume to substantially increase the number of SN Ia hosts that can be calibrated with Miras, it is possible that some of these galaxies can eventually become additional geometric anchors. Miras will be brightest in the $r, i,$ and $z$ bands, and most will be discovered in the $i$ band.  As previous optical studies have found, maximum-light Period-Luminosity Relations may be more useful and have smaller scatter than mean-light relations for these shorter wavelengths. 

The Nancy Grace Roman Observatory will have a similar sensitivity and resolution to Hubble, but will cover a much larger field-of-view. This will allow it to continue to study Miras at the distances of SN Ia hosts. In addition, it has a redder, $K$-band equivalent that may produce tighter P-L relations for Miras than the HST \textit{F160W} (H) band currently used in the distance ladder. Given its the large field of view, many Local Group Miras are also likely to fall into survey regions, potentially providing NIR data for previously-identified stars.



\begin{acknowledgement}
I would like to thank Eleonora di Valentino and Dillon Brout for inviting me to write this review chapter and for their hard work in assembling this textbook. I also thank Morgan MacLeod, John Menzies, Patricia A. Whitelock, Louise Breuval, and Andrea Sacchi for their helpful comments and discussions regarding this review. Additionally, I thank Morgan MacLeod for his help in the creation of Figure \ref{fig:HRDiagram}. 
\end{acknowledgement}







\clearpage

\bibliography{chapter}{}
\bibliographystyle{unsrt}


\end{document}